# Scale-free and small-world properties of earthquake network in Chile

Denisse Pastén[1], Sumiyoshi Abe[2,3], Víctor Muñoz[1] and Norikazu Suzuki[4]

[1]Departamento de Física, Facultad de Ciencias, Universidad de Chile,
Casilla 653, Santiago, Chile

[2]Department of Physical Engineering, Mie University, Mie 514-8507, Japan

[3]Institut Supérieur des Matériaux et Mécaniques Avancés, 44 F. A. Bartholdi,
72000 Le Mans, France

[4]College of Science and Technology, Nihon University, Chiba 274-8501, Japan

**Abstract.**  The properties of earthquake networks have been studied so far mainly for the seismic data sets taken from California, Japan and Iran, and features common in these regions have been reported in the literature. Here, an earthquake network is constructed and analyzed for the Chilean data to examine if the scale-free and small-world properties of the earthquake networks constructed in the other geographical regions can also be found in seismicity in Chile. It is shown that the result is affirmative: in all the regions both the exponent $\gamma$ of the power-law connectivity distribution and the clustering coefficient $C$ take the universal invariant values $\gamma \approx 1$ and $C \approx 0.85$, respectively, as the cell size becomes larger than a certain value, which is the scale of coarse graining needed for constructing earthquake network. An interpretation for this remarkable result is presented based on physical considerations.

# 1    Introduction

Dynamics governing seismicity is yet largely unknown, and accordingly the relevant research areas are necessarily empirical, owing their developments to case studies. This seems to be true even in the processes of establishing two celebrated classical laws known as the Omori law for temporal pattern of aftershocks and the Gutenberg-Richter law for the relation between frequency and magnitude.

In the absence of detailed knowledge about underlying dynamics, it is of central importance to clarify the pattern of correlations in order to extract information on features of the dynamics as a time series. Here, our primary interest is in the event-event correlations between successive earthquakes. As reported in the literature, both the spatial distance (Abe and Suzuki, 2003) and time interval (Corral, 2004; Abe and Suzuki, 2005) between two successive events strongly deviate from Poissonian statistics, supporting rationality of the concept of the event-event correlations. Accordingly, the following working hypothesis may be framed: two successive events are indivisibly correlated at least at the statistical level, no matter how distant they are. In this context, it is also noted that an earthquake can in fact trigger the next one more than 1000 km away (Steeples and Steeples, 1996). Thus, the correlation length can indeed be divergently large, indicating a similarity to critical phenomena.

Although seismicity is generically assumed to be a complex phenomenon, it is actually a nontrivial issue to characterize its complexity in a unified way.

The network approach offers a powerful tool to analyzing kinematical and dynamical structures of complex systems in a holistic manner (Buchanan, 2002, as a general reading). In a series of recent works (Abe and Suzuki, 2004, 2005, 2006a, 2009a), we have introduced and developed the concept of earthquake network in order to reveal the complexity of seismicity. In this approach, the event-event correlations are replaced by edges connecting vertices (for the details of constructing an earthquake network, see the next section). In those works, we have employed mainly the data sets taken from California and Japan. It was found that the networks constructed in these two regions are complex ones, which are scale-free (Barabási and Albert, 1999) and small-world (Watts and Strogatz, 1998). Close studies have shown (Abe and Suzuki, 2006b) that they are hierarchically organized (Ravasz E. and Barabási A.-L., 2003) and possess the property of assortative mixing (Newman, 2002). It has also turned out that time evolution of network characterizes main shock in a peculiar manner (Abe and Suzuki, 2007). In addition, it has been found that the scaling relation holds for the exponents of the power-law connectivity distribution and network spectral density (Abe and Suzuki, 2009b).

Thus, the network approach is expected to offer a novel possibility to deeper physical understandings of seismicity. Since the approach is inherently empirical in the sense that it is based on analyses of real data, it is essential to examine if the properties found from the data sets in California and Japan can also be recognized in other geographical regions. At the present early stage of the research, this point is of fundamental importance for establishing the universality of these properties of earthquake network.

Therefore, in this paper, we analyze, for the first time, the earthquake network constructed in a South American region, specifically in Chile. In particular, we focus our attention to the scale-free and small-world properties. We show that the earthquake network in Chile is indeed scale-free and small-world. In addition, we carefully study the dependencies of these two properties on a single parameter (i.e., the cell size of division needed for constructing a network: see Sect. 2) and compare them with those in other geographical regions. We see that the same behaviors as those in the other regions are in fact observed, adding a supporting evidence for the universality of these properties. Also, the scale of coarse graining (associated with the cell size) is determined for seismicity in Chile by the invariance principle. We show a remarkable fact that both the exponent $\gamma$ of the power-law connectivity distribution and the clustering coefficient $C$ take the universal invariant values $\gamma \approx 1$ and $C \approx 0.85$, respectively, as the cell size becomes larger than a certain value. We shall present a physical explanation about emergence of such invariance.

The present paper is organized as follows. In Sect. 2, we review the method of constructing an earthquake network. In Sect. 3, we discuss the scale-free and small-world properties of the earthquake network in Chile. In Sect. 4, we examine the dependence of the network on a parameter (i.e., the cell size) and compare it with those in other regions. The scale of coarse graining is determined there. Section 5 is devoted to concluding remarks.

## 2      Construction of earthquake network

The method of constructing an earthquake network (Abe and Suzuki, 2004) is as follows.

A geographical region under consideration is divided into cubic cells. A cell is regarded as a vertex of a network if earthquakes with any values of magnitude (above a certain detection threshold) occurred therein. Two successive events define an edge between two vertices, but if they occur in the same cell, then a tadpole (i.e., a self-loop) is attached to that vertex. These edges and tadpoles represent the event-event correlations (recall the working hypothesis mentioned in Sect. 1).

This simple procedure enables one to map a given seismic time series to a growing stochastic network, which is an earthquake network that we have been referring to.

Several comments on the construction are in order. Firstly, it contains a single parameter: the cell size $L$, which is the scale of coarse graining. That is, all earthquakes occurred in a given cell are identified and represented by the relevant vertex. Once a set of cells is fixed, then an earthquake network is unambiguously defined. Secondly, an earthquake network is a directed one in its nature. However, directedness does not bring any difficulties to statistical analysis of connectivity (or degree, i.e., the number of edges attached to the vertex under consideration) needed for examining the scale-free property, since by construction the in-degree and out-degree are identical for each vertex except the initial and final vertices in analysis, and so they need not be distinguished. Therefore, vertices except the initial and final ones have the even-number values of connectivity. Thirdly, a full directed earthquake network should be reduced to a simple undirected network, when its small-world property is examined (see Fig. 1). That is, tadpoles are removed and each multiple edge is replaced by a single edge.

To practically set up cells and identify a cell for each earthquake, here we employ the following procedure. Let $\theta_0$ and $\theta_{max}$ be the minimal and maximal values of latitude covered by a data set, respectively. Similarly, let $\phi_0$ and $\phi_{max}$ be the minimal and maximal values of longitude. Define $\theta_{av}$ as the sum of the values of latitude of all the events divided by the number of events contained in the analysis. The hypocenter of the $i$th event is denoted by $(\theta_i, \phi_i, z_i)$, where $\theta_i$, $\phi_i$ and $z_i$ are the values of latitude, longitude and depth, respectively. The north-south distance between $(\theta_0, \phi_0)$ and $(\theta_i, \phi_i)$ reads $d_i^{NS} = R \cdot (\theta_i - \theta_0)$, where $R \, (\cong 6370 \, \text{km})$ is the radius of the Earth. On the other hand, the east-west distance is given by $d_i^{EW} = R \cdot (\phi_i - \phi_0) \cdot \cos \theta_{av}$. In these expressions, all the angles should be described in the unit of radian. The depth is simply $d_i^D = z_i$. Now, starting from the point $(\theta_0, \phi_0, z_0 \equiv 0)$, divide the region into cubic cells with a given value of the cell size *L*. Then, the cell of the *i*th event can be identified by making use of $d_i^{NS}$, $d_i^{EW}$ and $d_i^D$.

Closing this section, we wish to emphasize the following point ascertained by our examinations. Although numerical values of characteristics of a network generically depend on how cells are set up, gross properties of a network do not change.

## 3  Scale-free and small world properties of earthquake network in Chile

The seismic data taken from Chile covers the period between 04:21:57.0 on October 2, 2000 and 18:31:57.3 on March 29, 2007, $29.01°S - 35.50°S$ latitude, $69.51°W - 73.95°W$ longitude with the maximal depth 293.30 km. The total number of earthquakes contained in the data is 17004. Most of them occurred in a shallow region: 90% of them are shallower than 116.30 km. Examination of the Gutenberg-Richter law shows that the law holds well for magnitude between 3.5 and 6. It seems therefore that detection of weak earthquakes was not performed.

We have constructed the earthquake network form this Chilean data following the procedure explained in Sect. 2 and analyzed its properties. In particular, we focus our attention to its scale-free and small-world properties.

In Fig. 2, we present a plot of the connectivity distribution (or, degree distribution) $P(k)$, which gives the probability of finding a vertex with $k$ edges. As can be seen, it is a power-law distribution

$$P(k) \sim \frac{1}{k^{\gamma}}, \qquad (1)$$

showing that the full earthquake network in Chile is in fact scale-free.

The scale-free nature indicates that there are quite a few hubs with large values of connectivity, which make the network heterogeneous. We have taken a close look at a hub and found that it corresponds to a main shock with M6.2 occurred at 13:41:25.6 on August 28, 2004, 35.17°S latitude, 70.53°W longitude and 5.00 km in depth. This shallow main shock was followed by a number of aftershocks. An interesting point is that, as observed in California and Japan, aftershocks tend to return to the locus of a main shock. This empirical fact makes a main shock play a role of a hub of a network.

Next, we discuss small-worldness of the reduced simple earthquake network in Chile. Here, the following two characteristic quantities are of central interest. One is the clustering coefficient and the other is the average path length (Watts and Strogatz, 1998), which are explained below.

The clustering coefficient $C$ of a simple network with $N$ vertices is defined by

$$C = \frac{1}{N}\sum_{i=1}^{N} c_i . \qquad (2)$$

$c_i$ appearing on the right-hand side is given by $c_i \equiv$ (number of edges of the $i$th vertex and its neighbors)$/[k_i(k_i-1)/2]]$ with $k_i$ being the connectivity of the $i$th vertex. Equivalently, it is calculated also as follows. Let $A$ be the symmetric adjacency matrix of the reduced simple network. Its element $(A)_{ij}$ is 1 (0), if the $i$th and $j$th vertices are linked (unlinked). Then, $c_i$ is also written as follows:

$$c_i = \frac{(A^3)_{ii}}{k_i(k_i-1)/2} . \qquad (3)$$

$c_i$ tells about the tendency that two neighboring vertices of the $i$th vertex are linked together. By definition, $C$ takes a value between 0 and 1. An important point is that $C$ of a small-world network is much larger than that of the corresponding random network (i.e., classical Erdös-Rényi random graph) given by

$$C_{random} = \frac{<k>}{N} \ll 1, \qquad (4)$$

where $<k>$ stands for the average value of connectivity of the random network. That is,

$$C \gg C_{random} . \qquad (5)$$

The other important characteristic quantity is the average path length $\bar{l}$, which quantifies that, for a pair of vertices, how many steps the shortest path linking them contains. A small-world network has a small value for it.

In Table 1, we pesent the results about these two characteristic quantities.

The above two results show that the reduced simple earthquake network in Chile is in fact of the small-world (as in California and Japan).

## 4   Disappearance of cell-size dependence

Quantities characterizing a network are usually dimensionless, as the connectivity distribution, clustering coefficient and average path length are. On the other hand, in the case of earthquake network, their numerical values depend on the cell size *L*, in general. This is the issue to be discussed in this section.

First of all, it is reasonable to make the cell size dimensionless. There are two possibilities for doing so: rescaling *L* by using the two or three dimensions of a geographical region under consideration. Let $L_{\text{LAT}}$, $L_{\text{LON}}$ and $L_{\text{DEP}}$ be the dimensions of the whole region in the directions of latitude, longitude and depth, respectively. From these, we construct the following two dimensionless quantities:

$$l_3 \equiv L / (L_{\text{LAT}} L_{\text{LON}} L_{\text{DEP}})^{1/3}, \tag{6}$$

$$l_2 \equiv L / (L_{LAT} L_{LON})^{1/2}. \tag{7}$$

At first glance, $l_3$ might seem more reasonable than $l_2$. However, it is not the case, in general, since as mentioned in the previous section 90% of the total events contained in the Chilean data are shallower than 116.30 km, whereas $L_{LAT} = 722.43$ km and $L_{LON} = 415.63$ km. Therefore, the region is quasi-two-dimensional. From the above-mentioned values as well as $L_{DEP} = 293.30$ km, we obtain $(L_{LAT} L_{LON} L_{DEP})^{1/3} = 444.91$ km and $(L_{LAT} L_{LON})^{1/2} = 547.96$ km.

Let us discuss the cell-size dependence of the exponent $\gamma$ in Eq. (1). In Fig. 3, we present plots of $\gamma$ of the scale-free earthquake networks in Chile. To calculate $\gamma$ from the data, we have employed the method of maximum-likelihood estimation for a power-law distribution. For comparison and observation of the universality of the result, we also present there those of the networks in California, Japan and Iran, details about which are found in the work (Abe and Suzuki, 2009c). As can be seen in Fig.3, the result is remarkable: $\gamma$ approaches a fixed value and the cell-size dependence disappears if the cell size becomes larger than a certain value $l_*$, which is the scale of coarse graining. In the case of $l_3$, the scale of coarse graining is roughly $l_{3*} \approx 0.05$, where as $l_{2*} \approx 0.04$. Both of them gives rise to $L_* \approx 22$ km in Chile. In addition, in all the four geographical regions, $\gamma$ takes the universal invariance value $\gamma \approx 1$.

The above result may be interpreted as follows. According to the method of network construction, there are two competitive factors. As the cell size increases, vertices get merged, yielding vertices with larger values of connectivity, while at the same time geographically neighboring vertices are absorbed each other, loosing the roles of hubs. The former decreases the value of $\gamma$, whereas the latter increases it. Disappearance of the cell-size dependence of $\gamma$ may be due to the balance between these two competitive effects.

Finally, let us discuss the cell-size dependence of the clustering coefficient *C*. In Fig. 4, we present its plots for the earthquake networks in Chile. Again, quite remarkably, *C* also takes the universal invariant value $C \approx 0.85$ as the cell size becomes larger than a certain value. Like before, for comparison and observation of the universality of the result, we also present in Fig. 4 those in California, Japan and Iran (Abe and Suzuki, 2009c).

This result may be explained as follows. As the cell size increases, the number of vertices decreases, and the network approaches a complete graph (i.e., a fully linked network) having the maximum value of the clustering coefficient ($C = 1$), but at the same time cells swallow triangular loops [recall the $A^3$ – nature of *C* in Eq. (3)] attached to them as they become larger. The former effect increases the value of *C*, whereas the latter decreases it. Disappearance of the cell-size dependence of *C* may be due to the balance between these two competitive mechanisms.

Unlike in the cases of California and Japan, unfortunately it does not seem to be possible to determine the scale of coarse graining in consistence with the one obtained from the analysis of $\gamma$. This is apparently due to smallness of the data size in Chile. The number of events contained in the data sets in California, Japan and Iran are 404106, 681547 and 22845, respectively (Abe and Suzuki, 2009c), whereas the Chilean data set contains only 17004 events. Clearly, a larger data set is desired in order to better recognize the universality of the characteristics of the Chilean earthquake network.

## 5    Concluding remarks

We have studied the scale-free and small-world properties of the earthquake network in Chile and compared them with those in California, Japan and Iran. We have observed that the network in Chile is in fact of the scale-free and small-world and its cell-size dependence qualitatively coincides with those in the other regions.

Based on the present results, we are firmly convinced that the discovered invariant values

$$\gamma \approx 1, \tag{8}$$

$$C \approx 0.85, \tag{9}$$

are universal and intrinsic in the seismicity of the Earth.

An additional remark is about incompleteness of a seismic data set. One might wonder if the incompleteness leads to change of the present results. Regarding this point, one should recall an important feature of a complex network: that is, it has a high degree of tolerance against "random attacks", i.e., random removals of vertices (Albert, Jeong and Barabási, 2000). Since incompleteness of a data set is not biased (i.e., not due to "intelligent attacks"), we can confidently assume robustness of the results presented here.

*Acknowledgements*.   The authors would like to thank Servicio Sismológico Nacional de Chile for providing them with the seismic data in Chile. D. P. is indebted to the scholarship of CONICyT No. 21070671 and the hospitality of Mie University extended to her. S. A. was supported in part by a Grant-in-Aid for Scientific Research from the Japan Society for the Promotion of Science.

581-586, 2004.

Abe, S. and Suzuki, N.: Scale-free statistics of time interval between successive earthquakes, Physica A 350, 588-596, 2005.

Abe, S. and Suzuki, N.: Complex-network description of seismicity, Nonlin. Processes Geophys. 13, 145-150, 2006a.

Abe, S. and Suzuki, N.: Complex earthquake networks: Hierarchical organization and assortative mixing, Phys. Rev. E 74, 026113, 2006b.

Abe, S. and Suzuki, N.: Dynamical evolution of clustering in complex network of earthquakes, Eur. Phys. J. B 59, 93-97, 2007.

Abe, S. and Suzuki, N.: Earthquake networks, complex, in Meyers, R. A. (Ed.) Encyclopedia of Complexity and Systems Science, 2530-2538, Springer, New York, 2009a.

Abe, S. and Suzuki, N.: Scaling relation for earthquake networks, Physica A 388, 2511-2514, 2009b.

Abe, S. and Suzuki, N.: Determination of the scale of coarse graining in earthquake networks, Europhys. Lett. 87, 48008, 2009c.

Albert R., Jeong H. and Barabási A.-L.: Error and attack tolerance of complex networks, Nature (London) 406, 378-382, 2000.

Barabási A.-L. and Albert R.: Emergence of scaling in random networks, Science 286, 509-512, 1999.

# Figure and Table Captions

**Fig. 1.** Schematic descriptions of earthquake network: (a) full directed network and (b) reduced undirected simple network. The vertices with the dotted lines indicate the initial and final events in analysis.

**Fig. 2.** The log-log plot of the connectivity distribution of the earthquake network in Chile for the cell size $20 \text{ km} \times 20 \text{ km} \times 20 \text{ km}$. The number of vertices is 1692. All quantities are dimensionless.

**Fig. 3.** Dependence of the exponent $\gamma$ on the dimensionless cell size: Chile ($\triangle$), California ($\bullet$), Japan ($\square$) and Iran ($\times$). (a) The plots with respect to $l_3$ and (b) the plots with respect to $l_2$. All quantities are dimensionless.

**Fig. 4.** Dependence of the clustering coefficient $C$ on the dimensionless cell size: Chile ($\triangle$), California ($\bullet$), Japan ($\square$) and Iran ($\times$). (a) The plots with respect to $l_3$ and (b) the plots with respect to $l_2$. All quantities are dimensionless.

**Table 1.** The values of the clustering coefficient $C$ and average path length $\bar{l}$. The cell size employed is $20 \text{ km} \times 20 \text{ km} \times 20 \text{ km}$, and the total number of vertices is 1692. For comparison, the value of the clustering coefficient of a corresponding random network, $C_{random}$, is also presented.

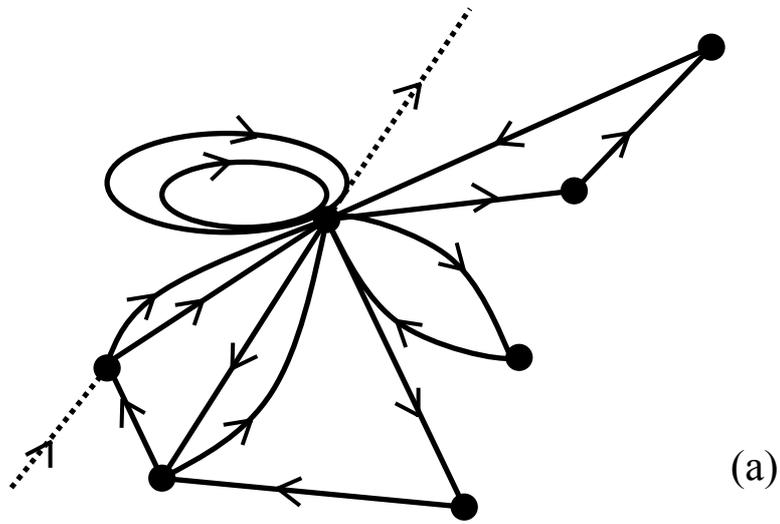

(a)

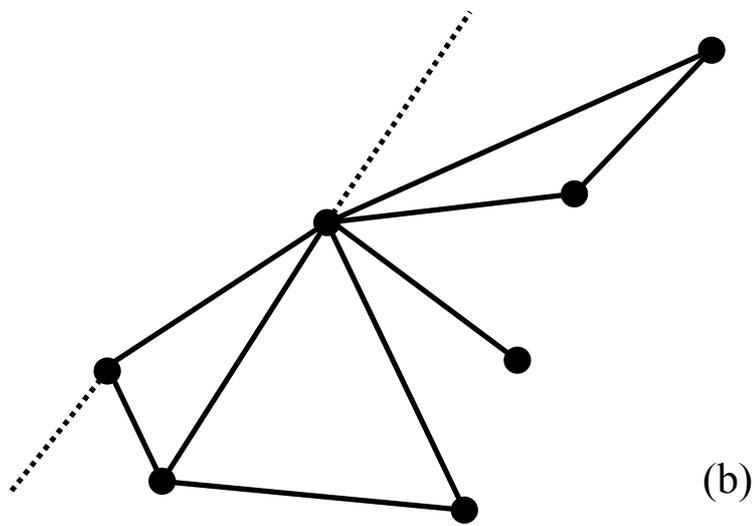

(b)

Fig. 1.

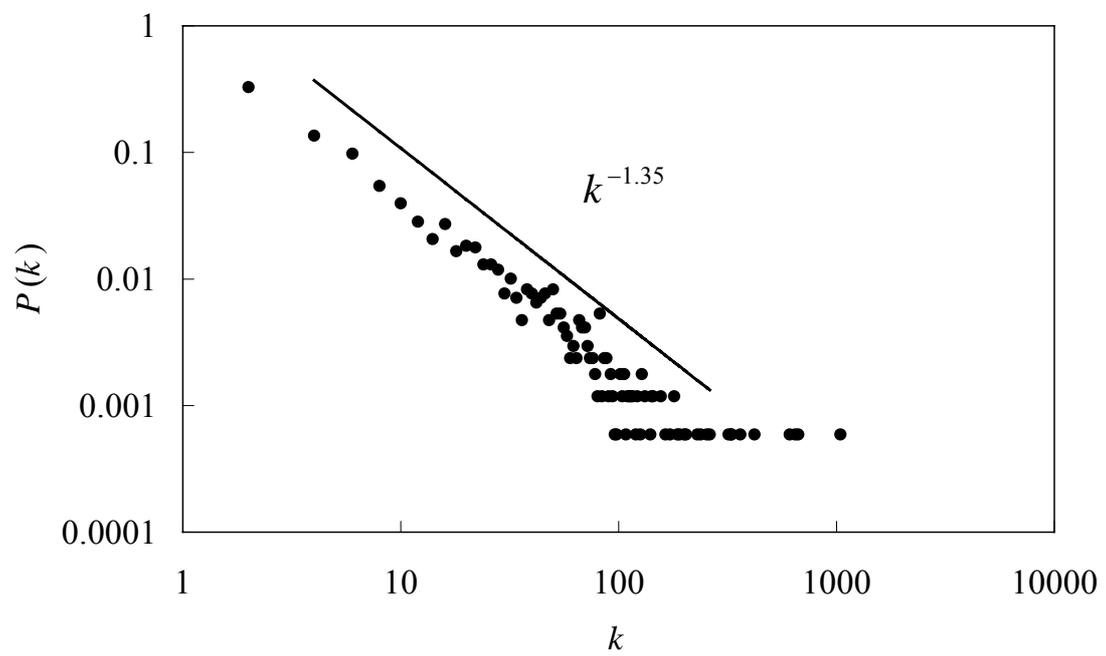

Fig. 2.

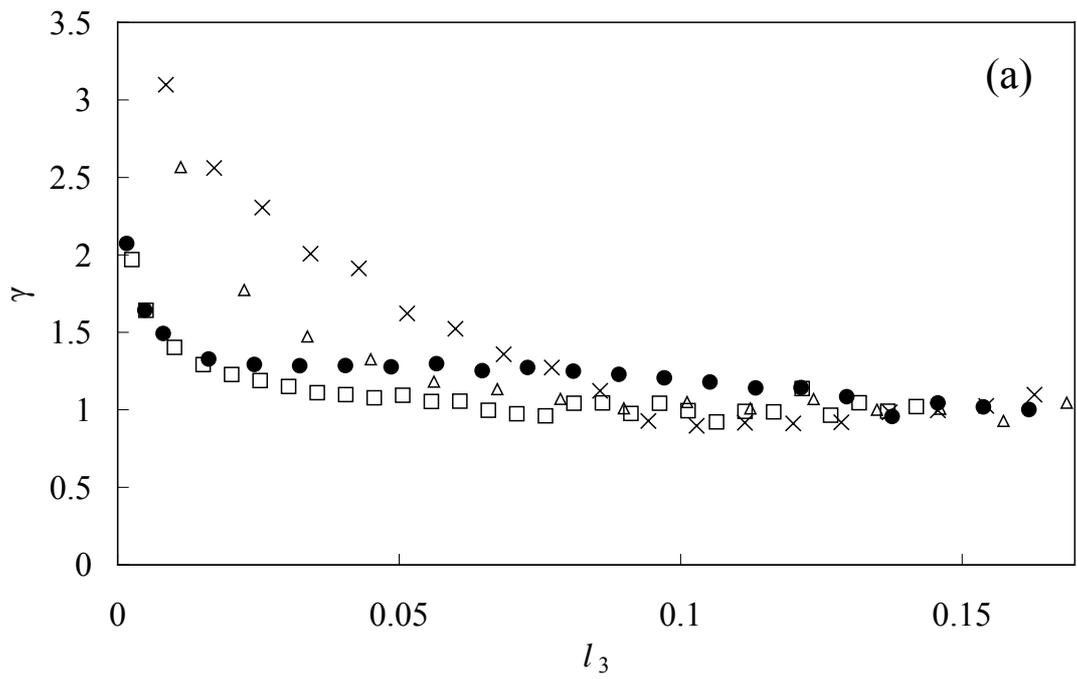

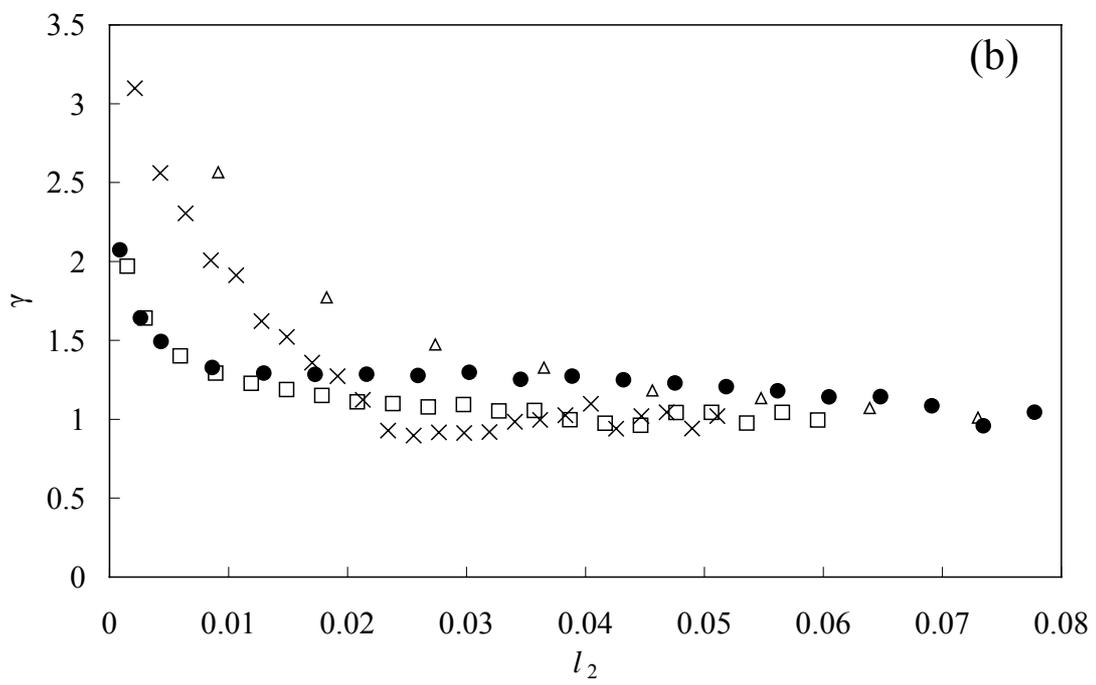

Fig. 3.

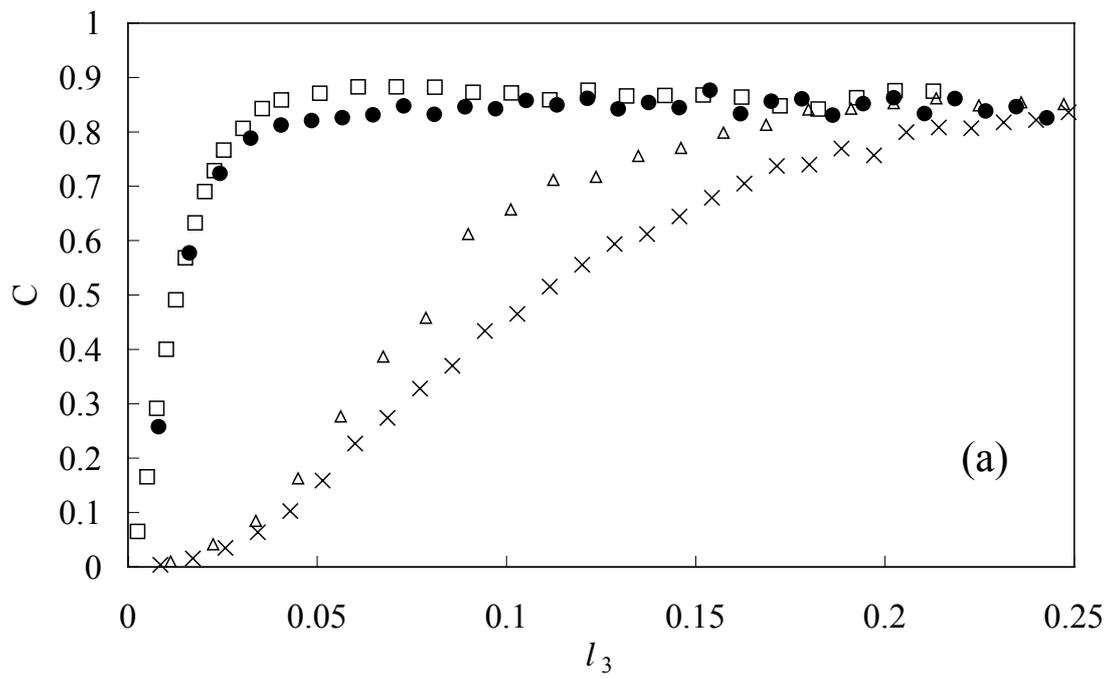

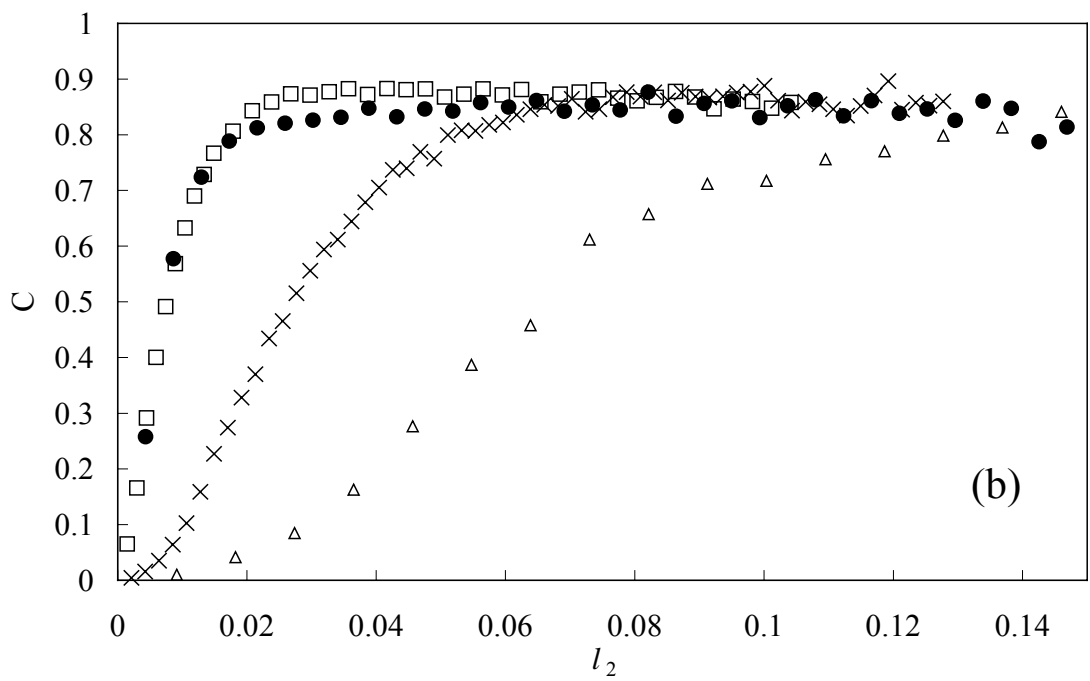

Fig. 4.

Table 1.

| $C$ | $C_{random}$ | $\bar{l}$ |
|---|---|---|
| 0.1629 | 0.0096 | 2.8783 |